\documentclass[sigconf, screen]{acmart}
\makeatletter
\renewcommand{\@authorfont}{\fontsize{11pt}{11pt}\selectfont\normalfont}
\renewcommand{\@affiliationfont}{\small\normalfont}
\makeatother
\usepackage[linesnumbered,ruled,vlined]{algorithm2e}
\SetAlCapFnt{\small}
\SetAlCapNameFnt{\small}
\makeatletter
\patchcmd{\@algocf@start}{-1.5em}{-5pt}{}{}
\makeatother
\usepackage{booktabs}
\usepackage{graphicx}           % rotatebox 
\usepackage{makecell}
\usepackage{multirow}
\usepackage[table, dvipsnames]{xcolor}
\usepackage[caption=false]{subfig}
\usepackage{float}              % H position
\usepackage{listings}
\usepackage{marvosym}           % Correspoding author e-mail
\usepackage{pifont}             % ding
\usepackage{xfrac}
\usepackage{amsmath,amsfonts}
\usepackage{framed}             % qtbox
\usepackage[most]{tcolorbox}    % theorem box
\usepackage{paralist}           % compactitem
\usepackage{changepage}         % adjustwidth
\usepackage{enumitem}
\usepackage{tikz}                   % Tikz
\usetikzlibrary{shapes.geometric}   % Step label
\usepackage{pgfplots} 
\pgfplotsset{compat=1.18}
\usepackage{kotex}              % Korean
\usepackage{lipsum}             % lipsum
\usepackage{paralist}           % Temp / compactenum, comactitem, comactdesc
\usepackage{comment}
\includecomment{WIP}

\definecolor{refgreen}{RGB}{0,204,0}
\newcommand*{\secref}[1]{\hyperref[#1]{\mbox{\textcolor{violet}{\S\ref*{#1}}}}\xspace}
\newcommand*{\appref}[1]{\hyperref[#1]{\textcolor{violet}{Appendix~\ref*{#1}}}\xspace}
\newcommand*{\figref}[1]{\hyperref[#1]{\textcolor{refgreen}{Figure~\ref*{#1}}}\xspace}
\newcommand*{\tabref}[1]{\hyperref[#1]{\textcolor{refgreen}{Table~\ref*{#1}}}\xspace}
\newcommand*{\algref}[1]{\hyperref[#1]{\textcolor{refgreen}{Algorithm~\ref*{#1}}}\xspace}
\newcommand*{\lstref}[1]{\hyperref[#1]{\textcolor{refgreen}{Listing~\ref*{#1}}}\xspace}
\renewcommand{\eqref}[1]{\hyperref[#1]{\textcolor{refgreen}{Equation~\ref*{#1}}}\xspace}
\newcommand*{\etal}{\textit{et~al.}\@\xspace}
\newcommand*{\ie}{\textit{i.\@e.\@,}\xspace}
\newcommand*{\eg}{\textit{e.\@g.\@,}\xspace}

\definecolor{Capproach}{RGB}{89,38,11}

\newcommand*{\CAA}{\hyperref[sec:method:CAA]{\textcolor{Capproach}{\textit{Core-Agnostic Architecture}}}\xspace}

\newcommand*{\AIS}{\hyperref[sec:method:AIS]{\textcolor{Capproach}{\textit{Adaptive Interference Suppression}}}\xspace}
\newcommand*{\DPPshort}{\hyperref[sec:method:DPP]{\textcolor{Capproach}{\textit{DPP}}}\xspace}
\newcommand*{\AISshort}{\hyperref[sec:method:AIS]{\textcolor{Capproach}{\textit{AIS}}}\xspace}
\newcommand*{\ICSHshort}{\hyperref[sec:method:ICSH]{\textcolor{Capproach}{\textit{ICSH}}}\xspace}
\definecolor{Cartifact}{RGB}{168,45,45}
\newcommand*{\CSM}{\textcolor{Cartifact}{\textit{Core Selection Map}}\xspace}
\newcommand*{\CSMshort}{\textcolor{Cartifact}{\ensuremath{\mathcal{M}_{cs}}}\xspace}
\newcommand*{\IMX}{\textcolor{Cartifact}{\textit{Interference Matrix}}\xspace}

\newcommand*{\BAOs}{\textcolor{Cartifact}{\textit{Baseline Adjustment Offsets}}\xspace}

\definecolor{Csbtype}{RGB}{0,102,102}
\newcommand*{\SBTYPE}[1]{\mbox{\textcolor{Csbtype}{\textit{Type-#1}}}\xspace}
\newcommand*{\freq}[1]{\ensuremath{f_{\text{#1}}}\xspace}
\newcommand*{\by}{\ensuremath{\times}\xspace}
\newcommand*{\map}[1]{\ensuremath{\mathcal{M}_{#1}}\xspace}

\newcommand*{\um}{\xspace\ensuremath{\mu m}\xspace}

\newcommand*{\ZynqTM}{Zynq\textsuperscript{TM}\xspace}
\newcommand*{\parHeading}[1]{\noindent{\textbf{#1.}}}

\newcommand*\BBCircle[1]{%
    \tikz[baseline=(char.base)]{%
        \node[shape=circle,fill,inner sep=1.2pt] (char) {\small\textcolor{white}{#1}};%
    }%
}%
\newcommand*\BWCircleX[1]{%
  \raisebox{0.19ex}{%
    \tikz[baseline=(char.base)]{%
        \node[shape=circle,fill=gray!20,inner sep=0.8pt,draw=black,line width=0.7pt] (char) {\small\textcolor{black}{\textbf{#1}}};%
    }%
  }%
}%
\definecolor{mycommentcolor}{RGB}{0,150,0}

\SetCommentSty{mycommentstyle}%
\SetKwInput{KwInput}{Input}%
\SetKwInput{KwOutput}{Output}%
\SetKw{ForEach}{for each}

\copyrightyear{2026}
\acmYear{2026}
\setcopyright{cc}
\setcctype{by}
\acmConference[DAC '26]{63rd ACM/IEEE Design Automation Conference}{July 26--29, 2026}{Long Beach, CA, USA}
\acmBooktitle{63rd ACM/IEEE Design Automation Conference (DAC '26), July 26--29, 2026, Long Beach, CA, USA}
\acmDOI{10.1145/3770743.3803928}
\acmISBN{979-8-4007-2254-7/2026/07}
\begin{document}
\title{Exploiting Per-Core Leakage: Electromagnetic Side-Channel Monitoring of Multicore Architectures}
\author{Daehyeon Bae}
\orcid{0000-0002-5523-6710}
\affiliation{
  \institution{Korea University}
  \city{Seoul}
  \country{South Korea}
}
\email{dh_bae@korea.ac.kr}

\author{Sujin Park}
\orcid{0009-0009-6594-732X}
\affiliation{
  \institution{Korea University}
  \state{Seoul}
  \country{South Korea}
}
\email{lemontrees33@korea.ac.kr}

\author{Insup Lee}
\orcid{0000-0002-9822-9860}
\affiliation{
  \institution{Korea University}
  \state{Seoul}
  \country{South Korea}
}
\email{islee94@korea.ac.kr}

\author{Younggiu Jung}
\orcid{0009-0001-6062-7417}
\affiliation{
  \institution{YM-NaeulTech.}
  \state{Incheon}
  \country{South Korea}
}
\email{youngq.jung@ym-naeultech.com}

\author{Kyeongsik Lee}
\orcid{0000-0002-1857-8105}
\affiliation{
  \institution{Agency for Defense Development}
  \state{Seoul}
  \country{South Korea}
}
\email{n0fate@add.re.kr}
\author{HeeSeok Kim}
\authornote{
    Corresponding author. This work was supported by the Agency for Defense Development by the Korean government~(No.~UG243060TD).
}
\orcid{0000-0001-8137-4810}
\affiliation{
  \institution{Korea University}
  \state{Sejong}
  \country{South Korea}
}
\email{80khs@korea.ac.kr}

\author{Seokhie Hong}
\orcid{0000-0001-7506-4023}
\affiliation{
  \institution{SmartM2M}
  \state{Busan}
  \country{South Korea}
}
\email{shhong@smartm2m.co.kr}
\begin{abstract}
Multicore processors are increasingly adopted in embedded systems to meet growing performance demands.
However, physical side-channel analysis of multicore architectures remains underexplored, as obtaining usable leakage is inherently challenging.
Consequently, side-channel security research on such systems has lagged far behind, leaving a critical security gap.
To address this gap, we reveal the electromagnetic leakage mechanisms in multicore architectures and, for the first time, demonstrate \textit{per-core leakage exploitation}, thereby enabling physical side-channel analysis for these systems.
As a practical extension, we present a non-intrusive \textit{side-channel monitoring} method that achieves per-core granularity.
To validate its feasibility and practicality, we implement a prototype on a heterogeneous SoC platform with an RF front-end, and evaluate on a commercial off-the-shelf quad-core embedded system, the Raspberry Pi 4B with ARM Cortex-A72 cores.
\vspace{-1ex}

\end{abstract}
\maketitle
\section{Introduction}
\label{sec:intro}

Advances in computing architectures have made physical side-channel analysis increasingly difficult. 
In particular, multicore architectures remain largely unexplored, as no effective solution has been proposed to address challenges arising from inter-core signal interference, signal heterogeneity, and dynamic core assignment driven by CPU scheduling policies and core affinity configurations.
Consequently, the field has fallen behind not only in cryptographic security evaluation but also in \textit{side-channel monitoring}.

Side-channel monitoring has recently gained significant attention as a promising security approach for observing power or electromagnetic (EM) signals externally from embedded systems, where on-device security solutions are inherently difficult to deploy~\cite{ieee-magz19:survey, hasp24:sok}.
Although multicore architectures are increasingly adopted in embedded systems to meet higher performance demands, research on such platforms remains limited, leaving a critical security gap.
While some argue that multicore systems can \textit{seemingly} host security solutions, side-channel monitoring in these systems remains both necessary and beneficial for the following reasons:
\begin{itemize}[label=\ding{93},itemsep=2pt,topsep=2pt,leftmargin=1.4em]
    \item \textbf{\textit{Persistent Resource Constraints.}}
    While increasing the number of cores generally enhances performance, overall system capability is also determined by the underlying microarchitecture of individual cores, memory and storage capacity, and I/O bandwidth.
    In fact, security solutions such as IDS, IPS, and antivirus software are resource-intensive---particularly in terms of memory and storage, often exceeding CPU requirements---making them difficult to deploy on multicore embedded systems.
    
    \item \textbf{\textit{Zero-Overhead.}} 
    The overhead introduced by on-device security solutions can undermine the consistency of latency and throughput, which is often unacceptable in mission-critical systems---especially in military and aerospace contexts~\cite{mtd20:mission-critical}. 
    Side-channel monitoring, however, avoids such overhead entirely, as it does not rely on the target device’s hardware resources.
    
    \item \textbf{\textit{Stealth Monitoring.}}
    Some sophisticated malware actively probes the system for signs of security components and initiates evasion strategies upon their discovery~\cite{acm-survey19:evasion, compupter-sec24:evasion}. 
    Side-channel monitoring, however, operates in a completely separate domain, making its execution entirely invisible to the target system.
\end{itemize}

\smallskip
\parHeading{Previous Approaches and Limitations}
% \parHeading{Related Work}
Only a few studies have \textit{explicitly} addressed physical side channels in multicore systems~\cite{tc23:MARCNNET, access25:multicore, usenix-sec15:multi-core-thermal}, excluding microarchitectural or cache-level work beyond the scope of this study~\cite{jhss18:multi-core-cache-survey, sp15:cross-core, ccs15:cross-core, sp22:cross-core}.
Yilmaz~\etal first explored side-channel monitoring on multicore systems, and their study remains the sole work in this domain~\cite{tc23:MARCNNET}.
Their approach, however, relies on \textit{aggregated} EM leakage---observed without considering the leakage mechanisms in multicore architectures---thereby leading to high complexity and unrealistic assumptions.
Navanesan~\etal investigated forensics using coarse-grained EM leakage in multicore systems, but their study was limited to cores under different clock domains~\cite{access25:multicore}.
Masti~\etal addressed inter-core thermal covert channels, which are not directly relevant to our work~\cite{usenix-sec15:multi-core-thermal}.

\smallskip
\parHeading{Our Approach and Contributions}
To achieve practical and low-complexity side-channel monitoring, we pursue an approach that leverages usable per-core leakage. 
In this context, we employ a \textit{fine-grained multi-probing} that directly measures non-aggregated EM leakage at individual cores.
While multi-probing has been shown to capture the leakage of individual circuits in FPGA-based cryptographic implementations~\cite{host18:multi-probe, asiacrypt17:location-x3}, extending this technique to multicore processors remains highly challenging due to the difficulty of identifying per-core leakage sources, inter-core interference, signal heterogeneity, and dynamic core assignment. 
In this work, building on our analysis of EM leakage mechanisms in multicore architectures, we address these inherent challenges and present the first method achieving per-core leakage exploitation. 
We summarize our contributions as follows:
\vspace{-0.5ex}
\begin{itemize}[label=\ding{93},leftmargin=1.4em]
    \item 
    We reveal the EM leakage mechanisms in multicore architectures and, for the first time, demonstrate per-core leakage exploitation, thereby enabling physical side-channel analysis for these systems.

    \item We analyze processor EM leakage spectra to characterize their patterns and implications, and based on these findings, propose an autoencoder-based side-channel monitoring method designed to be core-agnostic.

    \item To demonstrate feasibility and practicality, we implement a prototype device based on a heterogeneous SoC platform. 
    We further validate it on a commercial off-the-shelf quad-core system, Raspberry Pi 4B with ARM Cortex-A72 cores.
    
\end{itemize}

\smallskip
\parHeading{Organization} 
The remainder of this paper is organized as follows:
Section~\ref{sec:leakage} analyzes the EM leakage mechanisms in multicore architectures and analyzes the meaning of their spectral patterns. 
Section~\ref{sec:method} proposes a detailed method that enables side-channel monitoring on multicore architectures.
Section~\ref{sec:eval} presents the prototype implementation and experimental results.
Section \ref{sec:discus} discusses the limitations, and Section \ref{sec:conc} concludes the paper.

\section{EM Leakage in Multicore Architectures}
\label{sec:leakage}

We begin with a review of prior work on single-core EM leakage~(\secref{sec:leakage:prior-work}), followed by our own analysis and findings~(\secref{sec:leakage:sidebands}--\secref{sec:leakage:patterns}), and finally extend the analysis to multicore architectures~(\secref{sec:leakage:multicore}).

%% ┏━━━━━━━━━━━━━━━━━━━━━━━━━━━━━━━━━━━━━━━━━━━━━━━━━━━━━━━━━━━━━━━━━━━━━━━━━━┓
%% ┃ Amplitude Modulation in EM Side-Channel Leakage
%% ┗━━━━━━━━━━━━━━━━━━━━━━━━━━━━━━━━━━━━━━━━━━━━━━━━━━━━━━━━━━━━━━━━━━━━━━━━━━┛
\subsection{\texorpdfstring{Prior Work\,---\,Single-Core Leakage}{Prior Work~---~Single-Core Leakage}}
\label{sec:leakage:prior-work}

Previous studies have shown that EM side-channel leakage mainly arises from amplitude modulation~\cite{ches03:em, isca15:FASE}.
This modulation produces sideband patterns that are highly dependent on CPU core activity and have been widely exploited in the literature~\cite{micro16:spectral-profiling, tc20:REMOTE, usenix23:SPECTREM}.
These sidebands are known to reflect \textit{periodic activity} in the CPU core, with spectral peaks appearing at $\freq{c}\,{\pm}\,n{\cdot}\,\sfrac{1}{T}$, where \freq{c} denotes the clock frequency and $T$ is the period of the repeated activity, and $n\,{\in}\,\mathbb{N}$.
However, the mechanism behind sideband generation and its precise relationship with CPU activity remain not fully understood.

%% ┏━━━━━━━━━━━━━━━━━━━━━━━━━━━━━━━━━━━━━━━━━━━━━━━━━━━━━━━━━━━━━━━━━━━━━━━━━━┓
%% ┃ Fundamental Spectral Patterns
%% ┗━━━━━━━━━━━━━━━━━━━━━━━━━━━━━━━━━━━━━━━━━━━━━━━━━━━━━━━━━━━━━━━━━━━━━━━━━━┛
\subsection{Sideband Formation and Interpretation}
\label{sec:leakage:sidebands}

\parHeading{How Sidebands Are Formed}
A significant portion of EM side-channel leakage from CPU core originates in the \textit{on-chip power delivery network}~\cite{iccad17:simulation-em-res, asiaccs25:probeshooter}.
At each rising edge of the clock signal, CMOS circuits---comprising the CPU core---draw current as a result of switching activity, generating EM emissions at the clock frequency near the \textit{on-chip power rail}.
These emissions act as a carrier whose waveform depends on current draw.
When different instructions are executed in the CPU core, distinct circuit blocks become active, leading to changes in current on the power rail that modulate the carrier signal.
This serves as an amplitude modulator, leading to the formation of distinct sidebands.
This claim is supported by the equivalence between the sideband patterns measured at the decoupling capacitors---connected to the CPU core’s power rail---and those observed at the silicon die.

\smallskip
\parHeading{What Sidebands Reveal}
As established earlier, sidebands are fundamentally linked to the current drawn from the CPU core's power rail.
In this sense, the core’s power consumption is indirectly encoded in the sidebands.
Then, \textit{how exactly are the sideband patterns related to power consumption?}
As demonstrated by the simulations in~\secref{sec:leakage:patterns}, the non-sinusoidal nature of the carrier and the arbitrary modulating signal (power consumption) result in sideband patterns that are identical to the Fourier transform of the baseband signal---simply shifted and scaled by the carrier or its harmonics.
Thus, the sideband pattern corresponds to a frequency-shifted Fourier representation of the CPU core’s power consumption.

%% ┏━━━━━━━━━━━━━━━━━━━━━━━━━━━━━━━━━━━━━━━━━━━━━━━━━━━━━━━━━━━━━━━━━━━━━━━━━━┓
%% ┃ Fundamental Spectral Patterns
%% ┗━━━━━━━━━━━━━━━━━━━━━━━━━━━━━━━━━━━━━━━━━━━━━━━━━━━━━━━━━━━━━━━━━━━━━━━━━━┛
\subsection{Spectral Patterns}
\label{sec:leakage:patterns}

\pgfplotsset{
    type2_left/.style={
        width=4cm,
        height=2.4cm,
        xmin=0, xmax=200,
        ymin=0, ymax=15,
        minor x tick num=1,
        minor y tick num=1,
        ymajorgrids=true,     
        grid style=dashed,
        legend style={
            font=\scriptsize,
            column sep=3pt,
            cells={align=left},
            at={(0.97,0.70)},
            anchor=east
        },
        xticklabel style = {font=\footnotesize},
        yticklabel style = {
            /pgf/number format/fixed,
            font=\footnotesize
        },
        minor grid style={
            line width=0.3pt,
            draw=gray!80,
            dotted
        },
        ytick align=inside,
        xtick align=inside,
        title={Baseband},
        title style={font=\footnotesize, yshift=-0.2cm},
        xtick style={draw=none}
    }
}%
\pgfplotsset{
    type2_right/.style={
        width=5.2cm,
        height=2.4cm,
        xmin=800, xmax=1200,
        ymin=0, ymax=30,
        minor x tick num=1,
        minor y tick num=1,
        ymajorgrids=true,     
        grid style=dashed,
        ylabel={Mag.\\(1e-3)},
        ylabel style={font=\footnotesize, yshift=-4.95cm, align=center},
        legend style={
            font=\scriptsize,
            column sep=3pt,
            cells={align=left},
            at={(0.97,0.70)},
            anchor=east
        },
        xticklabel style = {font=\footnotesize},
        yticklabel style = {
            /pgf/number format/fixed,
            font=\footnotesize
        },
        minor grid style={
            line width=0.3pt,
            draw=gray!80,
            dotted
        },
        ytick align=inside,
        xtick align=inside,
        title={RF Band (Carrier Freq.)},
        title style={font=\footnotesize, yshift=-0.25cm},
        xtick style={draw=none}
    }
}%
\pgfplotsset{
    type3_left/.style={
        width=4cm,
        height=2.4cm,
        xmin=0, xmax=100,
        ymin=0, ymax=2.5,
        minor x tick num=1,
        minor y tick num=1,
        ymajorgrids=true,     
        grid style=dashed,
        legend style={
            font=\scriptsize,
            column sep=3pt,
            cells={align=left},
            at={(0.97,0.70)},
            anchor=east
        },
        xticklabel style = {font=\footnotesize},
        yticklabel style = {
            /pgf/number format/fixed,
            font=\footnotesize
        },
        minor grid style={
            line width=0.3pt,
            draw=gray!80,
            dotted
        },
        ytick align=inside,
        xtick align=inside,
    }
}%
\pgfplotsset{
    type3_right/.style={
        width=5.2cm,
        height=2.4cm,
        xmin=900, xmax=1100,
        ymin=0, ymax=5,
        minor x tick num=1,
        minor y tick num=1,
        ymajorgrids=true,     
        grid style=dashed,
        ylabel={Mag.\\(1e-3)},
        ylabel style={font=\footnotesize, yshift=-4.8cm, align=center},
        legend style={
            font=\scriptsize,
            column sep=3pt,
            cells={align=left},
            at={(0.97,0.70)},
            anchor=east
        },
        xticklabel style = {font=\footnotesize},
        yticklabel style = {
            /pgf/number format/fixed,
            font=\footnotesize
        },
        minor grid style={
            line width=0.3pt,
            draw=gray!80,
            dotted
        },
        ytick align=inside,
        xtick align=inside,
    }
}%
\pgfplotsset{
    type4_left/.style={
        width=4cm,
        height=2.4cm,
        xmin=0, xmax=100,
        ymin=0, ymax=1.5,
        minor x tick num=1,
        minor y tick num=1,
        ymajorgrids=true,     
        grid style=dashed,
        xlabel={Frequency [MHz]},
        xlabel style={font=\footnotesize, yshift=+.1cm, align=center},
        legend style={
            font=\scriptsize,
            column sep=3pt,
            cells={align=left},
            at={(0.97,0.70)},
            anchor=east
        },
        xticklabel style = {font=\footnotesize},
        yticklabel style = {
            /pgf/number format/fixed,
            font=\footnotesize
        },
        minor grid style={
            line width=0.3pt,
            draw=gray!80,
            dotted
        },
        ytick align=inside,
        xtick align=inside,
    }
}%
\pgfplotsset{
    type4_right/.style={
        width=5.2cm,
        height=2.4cm,
        xmin=900, xmax=1100,
        ymin=0, ymax=3,
        minor x tick num=1,
        minor y tick num=1,
        ymajorgrids=true,     
        grid style=dashed,
        xlabel={Frequency [MHz]},
        xlabel style={font=\footnotesize, yshift=+.1cm, align=center},
        ylabel={Mag.\\(1e-3)},
        ylabel style={font=\footnotesize, yshift=-4.8cm, align=center},
        legend style={
            font=\scriptsize,
            column sep=3pt,
            cells={align=left},
            at={(0.97,0.70)},
            anchor=east
        },
        xticklabel style = {font=\footnotesize},
        yticklabel style = {
            /pgf/number format/fixed,
            font=\footnotesize
        },
        minor grid style={
            line width=0.3pt,
            draw=gray!80,
            dotted
        },
        ytick align=inside,
        xtick align=inside,
    }
}%
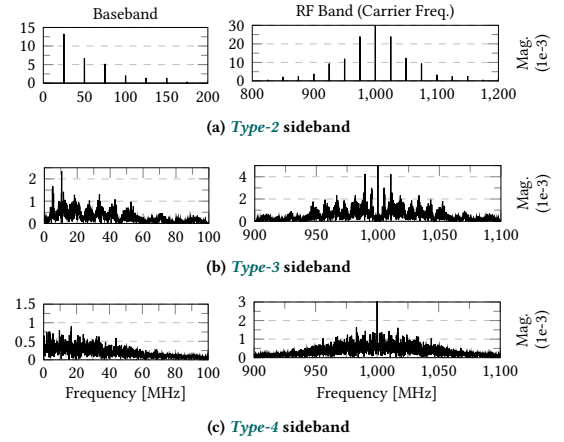
\begin{figure}[t]%
    \centering%
    \vspace{-0.3cm}
    \scalebox{0.9}{
        \subfloat[\centering \SBTYPE{2} sideband\label{fig:am-simulation:type2}]{%
            \centering
            \hspace{5pt}
            \begin{tikzpicture}
                \begin{axis}[type2_left]
                \addplot [color=black, line width=0.5pt]
                    table {figure/am-simulation/type2-bb.dat};
                \end{axis}
            \end{tikzpicture}
            \hspace{-7pt}
            \begin{tikzpicture}
                \begin{axis}[type2_right]
                \addplot [color=black, line width=0.5pt]
                    table {figure/am-simulation/type2-sb.dat};
                \end{axis}
            \end{tikzpicture}
        }%
    }
    \vspace{-0.2cm}
    \scalebox{0.9}{
        \subfloat[\centering \SBTYPE{3} sideband\label{fig:am-simulation:type3}]{%
            \centering
            \hspace{5pt}
            \begin{tikzpicture}
                \begin{axis}[type3_left]
                \addplot [color=black, line width=0.5pt]
                    table {figure/am-simulation/type3-bb.dat};
                \end{axis}
            \end{tikzpicture}
            \hspace{-7pt}
            \begin{tikzpicture}
                \hspace{4pt}
                \begin{axis}[type3_right]
                \addplot [color=black, line width=0.5pt]
                    table {figure/am-simulation/type3-sb.dat};
                \end{axis}
            \end{tikzpicture}
        }%
    }
    \vspace{-0.2cm}
    \scalebox{0.9}{
        \subfloat[\centering \SBTYPE{4} sideband\label{fig:am-simulation:type4}]{%
            \centering
            \hspace{5pt}
            \begin{tikzpicture}
                \hspace{-3pt}
                \begin{axis}[type4_left]
                \addplot [color=black, line width=0.5pt]
                    table {figure/am-simulation/type4-bb.dat};
                \end{axis}
            \end{tikzpicture}
            \hspace{-7pt}
            \begin{tikzpicture}
                \hspace{1pt}
                \begin{axis}[type4_right]
                \addplot [color=black, line width=0.5pt]
                    table {figure/am-simulation/type4-sb.dat};
                \end{axis}
            \end{tikzpicture}
        }%
    }
    \captionsetup{font=small}%
    % \vspace{-0.2cm}
    \caption{Atomic sideband patterns by activity type. These patterns were derived from AM simulations using a non-sinusoidal carrier and an arbitrary modulator to mimic EM leakage from CPU activity.}%
    \label{fig:am-simulation}%
    \vspace{-0.2cm}
\end{figure}%

In this subsection, we present \textit{atomic} sideband patterns that can emerge in EM leakage based on simulation results. 
The simulation models a non-sinusoidal carrier that is amplitude-modulated with a baseband signal mimicking CPU power consumption.

%% ---------------
\smallskip
\parHeading{\SBTYPE{1}---\,Idle State}
Some processors enter low-power modes by applying clock or power gating techniques when CPU operations are unnecessary, aiming to improve power efficiency. 
In such states, the core becomes inactive, suppressing amplitude modulation and thereby preventing sideband formation. 
Consequently, the resulting spectrum is clean, exhibiting either a single peak at the carrier frequency or no detectable carrier at all.

%% ---------------
\smallskip
\parHeading{\SBTYPE{2}---\,Strictly-Periodic Activity}
Repetitive activities are common in embedded systems due to their inherent nature~\cite{micro16:spectral-profiling}. 
When such repetition occurs in strictly identical cycles, the sideband pattern appears in the form of $\freq{c}\,{\pm}\,n{\cdot}\sfrac{1}{T}$, as introduced in~\secref{sec:leakage:prior-work}, with energy concentrated at a few sharp peaks—making them significantly higher than those of other types, as shown in~\figref{fig:am-simulation:type2}.

%% ---------------
\smallskip
\parHeading{\SBTYPE{3}---\,Jittered-Periodic Activity}
Repetitive behavior commonly exhibits slight jitter, especially when loops contain conditional branches. 
This case represents a variant of \SBTYPE{2}, where sharp peaks are spread into \textit{bumps}.
The center frequencies of the \textit{bumps} are slightly shifted in proportion to the accumulated jitter, as shown in~\figref{fig:am-simulation:type3}.

%% ---------------
\smallskip
\parHeading{\SBTYPE{4}---\,Irregular Activity}
Most software includes execution segments that appear (pseudo-) random and non-repetitive. 
In such cases, unlike \SBTYPE{2} and \SBTYPE{3}, the sideband does not exhibit peaks or bumps but instead forms a broad \textit{lobe} centered around the carrier frequency, as shown in~\figref{fig:am-simulation:type4}.
Due to their similar appearance, such patterns tend to lack distinct features, which makes them more difficult to model compared to other types.
For these reasons, this type remains largely unaddressed in the literature.

\smallskip
\parHeading{Compositional Space of Spectral Patterns}
All EM side-channel spectra manifest as superpositions (or instances) of selected patterns from \SBTYPE{1} to \SBTYPE{4}, resulting from their concatenation in the time domain.
Such concatenation inherently causes spectral leakage and slight deformation of the pattern, which are mitigated when a single pattern occupies a larger portion of the transform window.

%% ┏━━━━━━━━━━━━━━━━━━━━━━━━━━━━━━━━━━━━━━━━━━━━━━━━━━━━━━━━━━━━━━━━━━━━━━━━━━┓
%% ┃ Extending to Multicore
%% ┗━━━━━━━━━━━━━━━━━━━━━━━━━━━━━━━━━━━━━━━━━━━━━━━━━━━━━━━━━━━━━━━━━━━━━━━━━━┛
\subsection{Extending to Multicore Architectures}
\label{sec:leakage:multicore}

\begin{figure}[t]
    \centering%
    \includegraphics[width=0.89\linewidth]{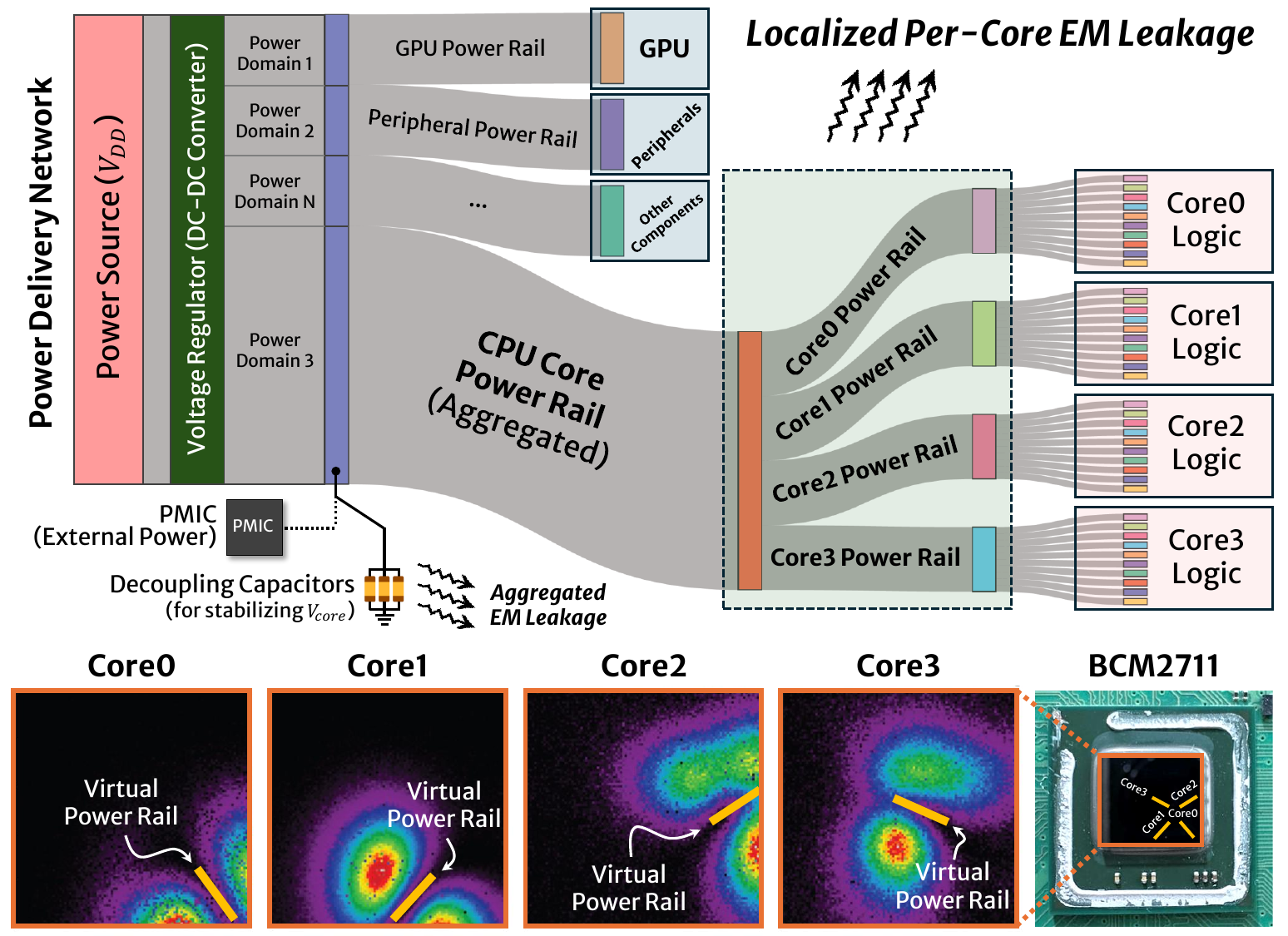}%
    \captionsetup{font=small}%1
    \vspace{-0.15cm}%
    \caption{Core-level EM leakage in a multicore processor. EM leakage is observed to be dominant and spatially localized near the virtual power rails of individual cores formed on the VDD plane. The corresponding on-chip power rails are indirectly identified through surface scanning with a horizontal coil (refer to \secref{sec:method:DPP}).}%
    \vspace{-0.2cm}%
    \label{fig:multi-core-leakage}%
\end{figure}%

Conceptually, multicore architectures consist of multiple cores fabricated on the same silicon die, each executing instruction-specific circuitry in the same manner. 
At the die level, each core draws current from its own on-chip power rail, resulting in localized EM leakage near the respective rail, as shown in~\figref{fig:multi-core-leakage}. 
However, no prior work has demonstrated a way to isolate such \textit{localized} leakages without inter-core interference. 
Consequently, only complex, superposed signals have been observed, hindering further progress in this area. 
If each leakage source could be spatially identified and inter-core interference mitigated, the single-core-level leakage characteristics could naturally extend to the multicore context, which is what this work enables.

\section{Proposed Method}
\label{sec:method}
\vspace{-0.8ex}

\begin{figure*}[t]
    \vspace{-0.2cm}%
    \centering%
    \includegraphics[width=0.95\linewidth]{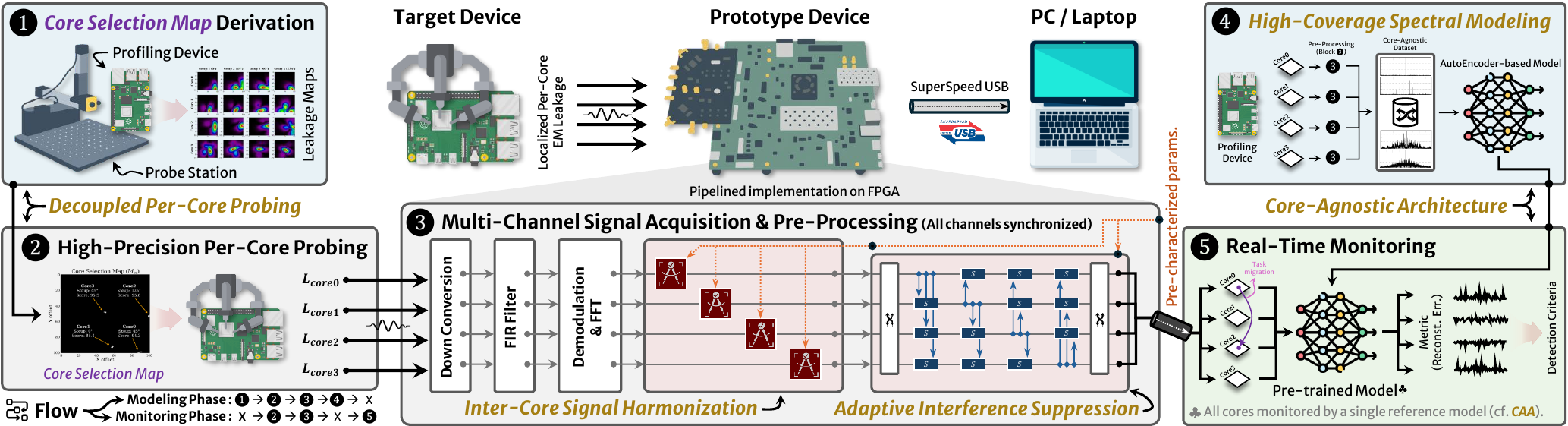}%
    \captionsetup{font=small}%
    \vspace{-0.2cm}%
    \caption{High-level representation of the proposed method. Block~\ding{184} operates on a fully pipelined and channel-synchronized device.}%
    \label{fig:overview}%
    \vspace{-1.5ex}
\end{figure*}%

A high-level overview of the proposed method is provided in \figref{fig:overview}. 
The specific workflow slightly differs depending on whether the modeling or monitoring phase is being conducted. The overall procedure consists of two primary phases as follows:
\vspace{-0.7ex}
\begin{itemize}[label=\ding{93},leftmargin=1.4em,itemsep=-2pt]
    \item Modeling phase : \BBCircle{1}$\rightarrow$\BBCircle{2}$\rightarrow$\BBCircle{3}$\rightarrow$\BBCircle{4}$\rightarrow$\BWCircleX{$\by$}
    \item Monitoring phase : \hspace{0.55ex}\BWCircleX{$\by$}$\rightarrow$\BBCircle{2}$\rightarrow$\BBCircle{3}$\rightarrow$\BWCircleX{$\by$}$\rightarrow$\BBCircle{5}
\end{itemize}

%% ┏━━━━━━━━━━━━━━━━━━━━━━━━━━━━━━━━━━━━━━━━━━━━━━━━━━━━━━━━━━━━━━━━━━━━━━━━━━┓
%% ┃ Threat model
%% ┗━━━━━━━━━━━━━━━━━━━━━━━━━━━━━━━━━━━━━━━━━━━━━━━━━━━━━━━━━━━━━━━━━━━━━━━━━━┛
\vspace{-1.8ex}
\subsection{Threat Model}
\label{sec:method:threat}
\vspace{-0.8ex}

We assume a scenario similar to conventional profiled side-channel analysis~\cite{ches02:template-attack, ches24:profiled-sca}, where normal behavior is modeled on a profiling device and the resulting model is subsequently used to monitor multiple target devices.
The profiling device must be capable of executing the target software under the same conditions as the target devices, but it requires no knowledge of the software itself.
The only privilege required on the profiling device is the ability to run small instruction sequences---\textit{gadgets}---with core affinity.

Our target devices are embedded systems designed for mission-specific tasks, excluding general-purpose platforms such as smartphones, tablets, and laptops. 
For example, our approach is well-suited for detecting abnormal behavior during inspection stages of military command and weapon systems, including communication platforms, missile systems, and unmanned aerial vehicles~(UAVs). 
Since we assume a fixed clock frequency, we do not consider dynamic voltage and frequency scaling (DVFS) in this work.

%% ┏━━━━━━━━━━━━━━━━━━━━━━━━━━━━━━━━━━━━━━━━━━━━━━━━━━━━━━━━━━━━━━━━━━━━━━━━━━┓
%% ┃ Core-Isolated Probing
%% ┗━━━━━━━━━━━━━━━━━━━━━━━━━━━━━━━━━━━━━━━━━━━━━━━━━━━━━━━━━━━━━━━━━━━━━━━━━━┛
\vspace{-1.5ex}
\subsection{Decoupled Per-Core Probing (DPP)}
\label{sec:method:DPP}
\vspace{-0.8ex}

In general, leakage sources on a chip are identified through leakage localization, which yields a leakage map~\cite{asiaccs25:probeshooter, access20:SCNIFFER, temc22:ANOVA, iet24:pattern-clustering}.
However, the leakage map on the chip surface is not invariant; it varies depending on probing factors---such as coil orientation, diameter, angle, and height---which also alter inter-core interference characteristics.
Accordingly, the key idea is to acquire leakage maps under diverse configurations and to identify, for each core, the probing location and conditions that minimize interference from neighboring cores.

To this end, multiple leakage maps are first acquired, from which a \CSM (\CSMshort) is derived using~\algref{alg:dpp}.
The most straightforward and effective way to obtain diverse leakage maps is to vary the rotation angle of the coil with respect to the chip’s origin.
For the acquisition of leakage maps, we employ an instruction snippet---referred to as a gadget (\eg see~\figref{fig:gadget})---which is designed to generate an artificial \SBTYPE{2} sideband leakage.
Since the sideband peaks directly reflect the underlying leakage as discussed in~\secref{sec:leakage:sidebands}, we can indirectly obtain the leakage map by measuring their power spectral density.
The resulting \CSMshort represents, for each core, the locations and conditions under which interference from other cores is minimized, thereby enabling fine-grained per-core probing.
Although \CSMshort-guided probing significantly reduces interference from neighboring cores, it does not completely eliminate it.
Nevertheless, the residual interference can be effectively mitigated by the \AIS (\secref{sec:method:AIS}), which leverages sideband characteristics for adaptive suppression.
Note that the \CSMshort needs to be derived only once per device type and remains valid across all devices equipped with the same chip.

\newcommand*{\CID}{\textit{core\_id}\xspace}
\newcommand*{\SID}{\textit{setup\_id}\xspace}
\newcommand*{\NUMOF}[1]{$N_{#1}$}
\newcommand*{\THR}[1]{$T_{#1}$}
\newcommand*{\TO}{\textnormal{{to}}\xspace}
\newcommand*{\NOT}{\textnormal{{not}}\xspace}
\newcommand*{\MAP}[1]{\ensuremath{\mathcal{M}_{#1}}}
\newcommand*{\INF}{\ensuremath{I_{upper}}}
\newcommand*{\SIG}{\ensuremath{S_{lower}}}
\newcommand*{\PTS}[1]{\ensuremath{\mathcal{P}_{#1}}}
\begin{algorithm}[t]%
\phantomsection
\footnotesize
\DontPrintSemicolon%
\KwInput{Leakage Maps~(\map{L}), Upper bound of inter-core interference~(${\alpha_{1}}{\%}$), Lower bound of desired signal~(${\alpha_{2}}{\%}$), Minimum valid pts per cluster~($\beta$), Scoring weight~($w$)}
\KwOutput{\CSM (\CSMshort)}
\vspace{1mm}
% ---------- Algorithm START ----------
$\mathcal{C} \gets$ [] \tcp*{Result candidates}
\For{{\normalfont\textbf{all}} (\SID, \CID)~{\normalfont\textbf{in}}~\map{L}} 
{
        \map{t} $\gets$ \MAP{L}$[\SID][\CID]$ \tcp*{Map of target core}
        \map{o} $\gets$ \MAP{L}$[\SID][\neq\CID]$ \tcp*{Maps of other cores}
        \For{\INF $\gets$ 0 \TO $\alpha_{1}$; \SIG $\gets$ $\alpha_{2}$ \TO 100} 
        {
                \PTS{signal} $\gets$ All points such that \map{t} > max(\map{t})\,$\cdot$\,\SIG \;
                \PTS{clean} $\gets$ All points such that \map{o} < max(\map{o})\,$\cdot$\,\INF \;
                Extract common points from \PTS{signal} and \PTS{clean}\;
                \If{\textnormal{DBSCAN\_validate\_cluster(\PTS{r}, $\beta$)}}
                {
                    \textit{score} $\gets$ $\SIG-w\,\cdot\,\INF$\;
                    Append (\CID, \SID, \PTS{r}, \textit{score}) to $\mathcal{C}$\;
                }
            % }
        }
    % }
}
\Return{{\normalfont argmax-}\textit{score}~{\normalfont tuple for each core in }$\mathcal{C}$}
% ---------- Algorithm END ----------
\caption{\CSM derivation.}\label{alg:dpp}
\end{algorithm}%
\lstdefinelanguage{gadget-bcm}{
  sensitive = true,
  keywords={gadget},
  otherkeywords={% Operators
    >, <, ==
  },
  keywords = [2]{mov,udiv,b,movk,},
  keywordstyle=\color{orange},
  keywordstyle=[2]\color{blue},
  showstringspaces=false,
  breaklines=true,
  frame=bottom,
  comment=[l]{;},
  morecomment=[s]{/*}{*/},
  commentstyle=\color{purple}\ttfamily,
  stringstyle=\color{red}\ttfamily,
  morestring=[b]',
  morestring=[b]"
}%
\begin{figure}[ht]%
    \vspace{-1.5ex}
    \center
    \begin{minipage}{0.5\linewidth}%
        \begin{lstlisting}[language=gadget-bcm,frame=tlrb,basicstyle=\scriptsize]
mov x1, #0xffffffffffffffff
mov x2, #0xff11            
gadget:                    
udiv x0, x1, x2            
b gadget           
        \end{lstlisting}%
    \end{minipage}%
    \captionsetup{font=small}%
    \vspace{-0.3cm}%
    \caption{Gadget for BCM2711 designed to induce artificial spectral leakage. The gadget repeats every 17 cycles, creating artificial peaks at $\freq{clk}\,{\pm}n{\cdot}\,\sfrac{\freq{clk}}{17}$, where \freq{clk} denotes the clock frequency and $n\,{\in}\,\mathbb{N}$.}%
    \label{fig:gadget}%
    \vspace{-0.2cm}
\end{figure}%

%% ┏━━━━━━━━━━━━━━━━━━━━━━━━━━━━━━━━━━━━━━━━━━━━━━━━━━━━━━━━━━━━━━━━━━━━━━━━━━┓
%% ┃ Inter-Core Harmonization
%% ┗━━━━━━━━━━━━━━━━━━━━━━━━━━━━━━━━━━━━━━━━━━━━━━━━━━━━━━━━━━━━━━━━━━━━━━━━━━┛
\vspace{-1.5ex}
\subsection{Inter-Core Signal Harmonization (ICSH)}
\label{sec:method:ICSH}
\vspace{-0.8ex}

\ICSHshort minimizes signal heterogeneity as a prerequisite for both \AIS~(\secref{sec:method:AIS}) and \CAA~(\secref{sec:method:CAA}).
The key idea of this approach is to harmonize signal levels and spectral baselines (noise floors) by adjusting other channels to match a designated reference channel (core).
To this end, \algref{alg:icsh} is used to derive the optimal gains that produce similar signal levels, as well as the \BAOs that characterize deviations in the noise floor.
During this process, a \textit{gadget} is also employed to artificially generate \SBTYPE{2} sidebands serving as a reference.
Subsequently, the signal levels and baselines of the spectra are harmonized in real time on the prototype device.

\begin{algorithm}[t]%
\phantomsection
\footnotesize
% \small
\DontPrintSemicolon%
\KwInput{Reference core~($c_{ref}$), Other cores ($\mathcal{C}_{others}$), Number of spectra for averaging~($N_{spectra}$)}
\KwOutput{Gains~($\mathcal{G}$), \BAOs~($\mathcal{O}$)}
\vspace{1mm}
% ---------- Algorithm START ----------
Run \textit{gadget} on $c_{ref}$\;
$S_{ref} \gets$ Sweep and averaging $N_{spectra}$ spectra \tcp*{Reference spectrum}
Stop \textit{gadget} on $c_{ref}$\;
$\eta_{ref} \gets$ Calculate baseline level\;
\For{{\normalfont\textbf{each}} $c$ \textnormal{in} $\mathcal{C}_{others}$}
{
    Run \textit{gadget} on $c$\;
    $\mathcal{D} \gets []$\tcp*{Result candidates}
    \For{{\normalfont\textbf{each}} g \textnormal{in available gains}}
    {
        Set gain $g$ to channel $c$\;
        $S_{g} \gets$ Sweep and average $N_{spectra}$ spectra\;
        $\eta_{g} \gets$ Calculate baseline level\;
        $d$ $\gets$ Euclidean dist. between $S_{ref}-\eta_{ref}$ and $S_{g}-\eta_{g}$\;
        Append ($d$, $g$, $\eta_{g}$) to $D$\;
    }
    Stop \textit{gadget} on $c$\;
    Append \textit{g} and \textit{$\eta_{g}$} of argmin \textit{d} in $\mathcal{D}$ to $\mathcal{G}$ and $\mathcal{O}$, respectively\;
}
\Return{$\mathcal{G}, \mathcal{O}$}
% ---------- Algorithm END ----------
\caption{\BAOs calculation.}\label{alg:icsh}
\end{algorithm}%

%% ┏━━━━━━━━━━━━━━━━━━━━━━━━━━━━━━━━━━━━━━━━━━━━━━━━━━━━━━━━━━━━━━━━━━━━━━━━━━┓
%% ┃ Adaptive Inter-Core Interference Suppression
%% ┗━━━━━━━━━━━━━━━━━━━━━━━━━━━━━━━━━━━━━━━━━━━━━━━━━━━━━━━━━━━━━━━━━━━━━━━━━━┛
\subsection{Adaptive Interference Suppression (AIS)}
\label{sec:method:AIS}

\begin{figure*}[t]%
    \vspace{-0.5cm}%
    \centering%
    \subfloat[\centering Experimental setup with the prototype device\label{fig:setup:all}]{%
        \includegraphics[height=4.52cm]{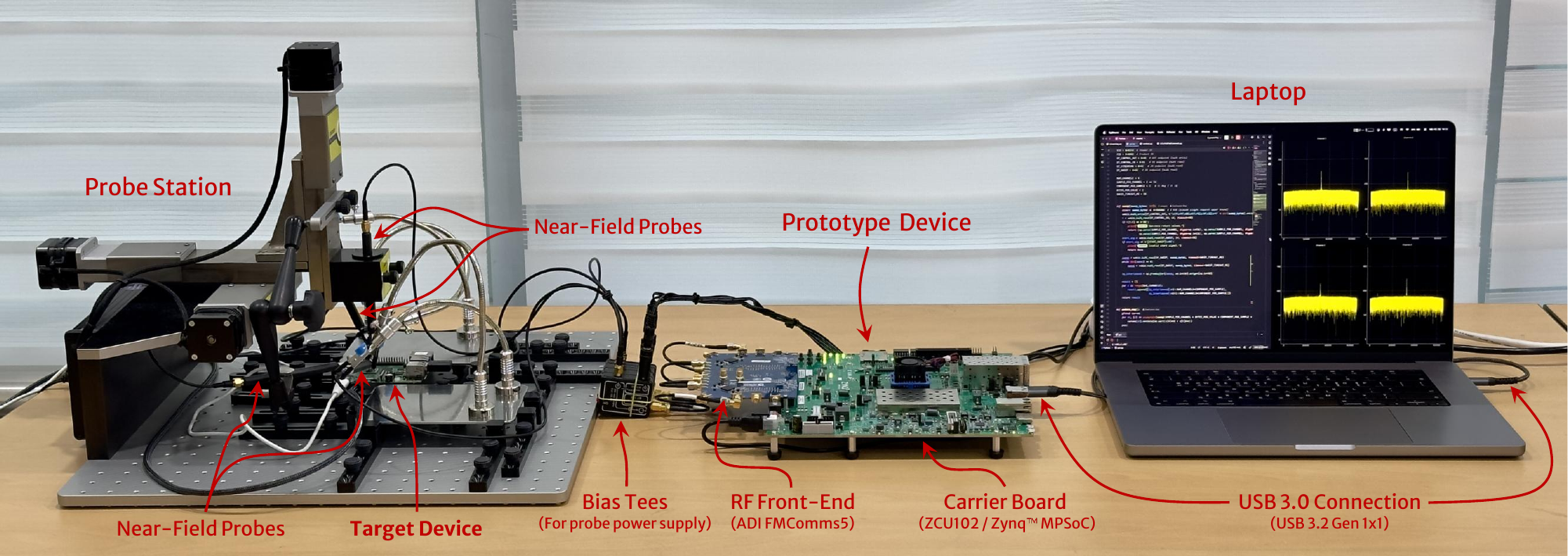}%
    }%
    \hspace{10pt}%
    \subfloat[\centering Per-core probing\label{fig:setup:probing}]{%
        \includegraphics[height=4.52cm]{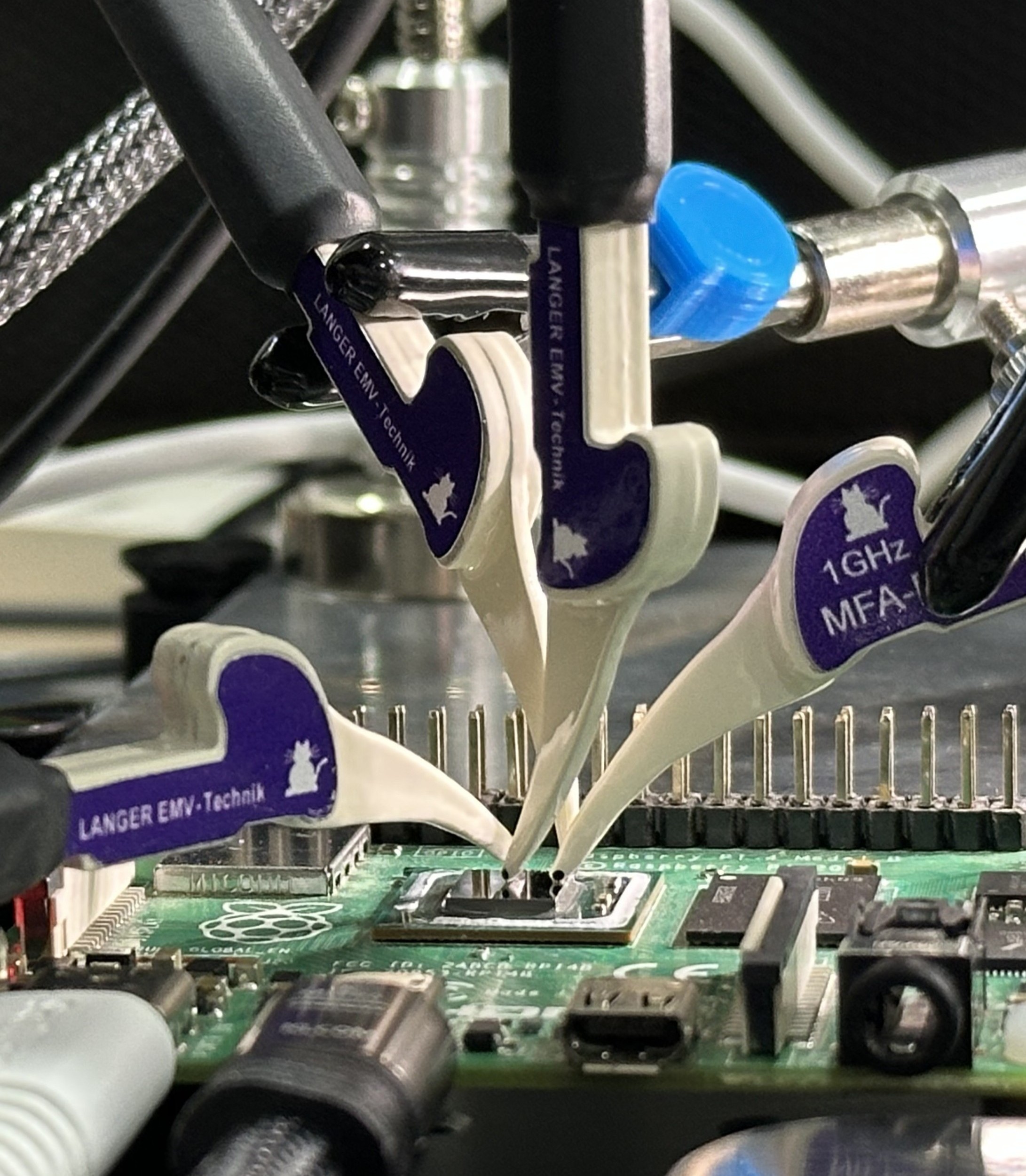}%
    }%
    \captionsetup{font=small}%
    \vspace{-0.25cm}
    \caption{Experimental setup of the proposed method. The probe station is used for \CSMshort derivation.}%
    \label{fig:setup}%
    \vspace{-1ex}
\end{figure*}%

The degree of interference from neighboring cores varies significantly depending on their activity---\ie the type of sidebands~(\secref{sec:leakage:patterns}).
Among these, we focus on \SBTYPE{2} sidebands, which constitute the dominant interference component, as supported by Parseval’s theorem.
That is, although different CPU activities produce different EM leakage patterns, the overall energy (L2 norm) remains largely consistent across types---both in time and frequency domains.
\SBTYPE{2} activity, however, concentrates energy into a few discrete peaks producing strong interference that is clearly visible to neighboring cores.
In contrast, other types disperse energy more broadly, causing interference sidebands to be buried in the noise floor.

Per-core signals can be simultaneously acquired with \textit{perfect synchronization}, enabled by the prototype device (refer to~\secref{sec:eval:prototype}). Accordingly, the key idea is to \textit{adaptively} estimate and suppress inter-core interference on a per-frame basis.
To this end, we first derive an \IMX that characterizes the degree of leakage coupling between all cores using~\algref{alg:ais}.
A gadget is also used in this step, as in \ICSHshort, and the same one can be reused.
Each value in the \IMX denotes the ratio of sideband magnitude between neighboring-core activity and self-core activity, ranging from 0 (none) to 1 (strong interference).
\algref{alg:ais2} then performs in-line interference suppression on the prototype device based on the derived \IMX.
To ensure that only genuine interference is removed, the procedure must be applied in the order of cores that are least affected by others.

%% ┏━━━━━━━━━━━━━━━━━━━━━━━━━━━━━━━━━━━━━━━━━━━━━━━━━━━━━━━━━━━━━━━━━━━━━━━━━━┓
%% ┃ Model Structure
%% ┗━━━━━━━━━━━━━━━━━━━━━━━━━━━━━━━━━━━━━━━━━━━━━━━━━━━━━━━━━━━━━━━━━━━━━━━━━━┛
\subsection{High-Coverage Spectral Modeling}
\label{sec:method:HCSM}

To model the sidebands of the EM leakage arising from the primitives introduced in \secref{sec:leakage:sidebands}, we adopt a signal-centric approach that avoids intermediate abstractions---such as state information~\cite{tc20:REMOTE, tc23:MARCNNET}---often employed in prior work. 
That is, we learn only the inherent patterns of the signal and detect anomalies as deviations from this learned behavior.
To this end, we adopt an unsupervised vanilla autoencoder as the backbone, owing to its simplicity, high throughput, and proven effectiveness in anomaly detection~\cite{arxiv:ae-survey}. 
We train the autoencoder to minimize the reconstruction error of individual FFT frames, where frames with high reconstruction errors---reflecting unlearned patterns---are identified as anomalies during inference. 
Although state-of-the-art architectures such as Transformers~\cite{neuroips17:transformer} offer greater modeling capacity, their inference latency and resource demands render them impractical.

\begin{algorithm}[t]%
\phantomsection
\footnotesize
% \small
\DontPrintSemicolon%
\KwInput{Number of spectra for averaging~($N_{spectra}$), Available CPU cores~($\mathcal{C}_{all}$), Gains from \ICSHshort~($\mathcal{G}$)}
\KwOutput{\IMX~($M$)}
\vspace{1mm}
% ---------- Algorithm START ----------
Set per-channel gains from $\mathcal{G}$\;
\For{{\normalfont\textbf{each}} $c_{intf}$ \textnormal{in} $\mathcal{C}_{all}$}
{
    Run \textit{gadget} on $c_{intf}$\;
    $\mathcal{S}_{intf} \gets$ Sweep and average $N_{spectra}$ spectra on $c_{intf}$\;
    $\rho_{intf} \gets$ Calculate signal level of $\mathcal{S}_{intf}$\;
    \For{{\normalfont\textbf{each}} $c_{v}$ \textnormal{in} $\mathcal{C}_{all}$ \textnormal{excluding} $c_{intf}$}
    {
        $\mathcal{S}_{v} \gets$ Sweep and average $N_{spectra}$ spectra on $c_{v}$\;
        $\rho_{v} \gets$ Calculate signal level of $\mathcal{S}_{v}$\;
        $M[c_{intf}][c_{v}] \gets \rho_{intf} / \rho_{v}$\tcp*{Interference ratio}
    }
    Stop \textit{gadget} on $c_{intf}$\;
}
\Return{$M$}
% ---------- Algorithm END ----------
\caption{\IMX generation.}\label{alg:ais}
\end{algorithm}%
\begin{algorithm}[t]%
\phantomsection
\footnotesize
% \small
\DontPrintSemicolon%
\KwInput{Interfering core spectrum~($S_{intf}$), Neighboring core spectra~($\mathcal{S}_{victims}$), \IMX~($M$), Adjustment factor~($\gamma$)}
\vspace{1mm}
% ---------- Algorithm START ----------
$i_{max} \gets$ max($S_{intf}$)\tcp*{Maximum interference magnitude}
\For{{\normalfont\textbf{each}} $S_{v}$ \textnormal{in} $\mathcal{S}_{victims}$}
{
    $\iota \gets$ Corresponding interference ratio from the $M$\;
    $th_{sup} \gets i_{max} \cdot (1 - \iota~+~\gamma)$\tcp*{Threshold}
    $S_{v} \gets$ max(0, $S_{v}-th_{sup}$)\tcp*{Update $S_{v}$ (intf. suppression)}
}
% ---------- Algorithm END ----------
\caption{In-line interference suppression.}\label{alg:ais2}
\end{algorithm}%

%% ┏━━━━━━━━━━━━━━━━━━━━━━━━━━━━━━━━━━━━━━━━━━━━━━━━━━━━━━━━━━━━━━━━━━━━━━━━━━┓
%% ┃ Multicore-Aware Architecture
%% ┗━━━━━━━━━━━━━━━━━━━━━━━━━━━━━━━━━━━━━━━━━━━━━━━━━━━━━━━━━━━━━━━━━━━━━━━━━━┛
\subsection{Core-Agnostic Architecture}
\label{sec:method:CAA}

This approach is an architecture for modeling and monitoring per-core signals that addresses dynamic core assignment and pursues structural simplicity.
The main idea is to eliminate the structural boundaries across cores, as illustrated in~\figref{fig:overview} (block~\BBCircle{4} and~\BBCircle{5}), where a single reference model is used regardless of the number of cores.
However, given the heterogeneous nature of per-core signals and the presence of inter-channel measurement variation, \ICSHshort must be used in conjunction with this architecture.
This design ensures that task assignment and migration have no impact, and that the load-balancing policy no longer needs to be considered.

\section{Implementation and Evaluation}
\label{sec:eval}

\subsection{Prototype Device}
\label{sec:eval:prototype}

The proposed method requires proportionally more RF channels and their associated resource demands---such as high data rates and throughput---which may cast doubt on its feasibility.
In addition, the \AISshort requires perfect synchronization of signals across channels (cores).
To address these concerns, we implement a prototype device for synchronized \textit{multi-channel} signal acquisition and pre-processing.
The prototype device consists of AMD’s ZCU102, a heterogeneous SoC platform based on the \ZynqTM UltraScale+ MPSoC~\cite{zcu102}, and Analog Devices’ FMComms5, a four-channel RF front-end~\cite{ad9361}.
The core logic is implemented on the FPGA in a pipelined manner, as illustrated in~\figref{fig:overview} (block~\BBCircle{3}).
We also develop bare-metal firmware for a Cortex-A53 core in the Zynq MPSoC, handling RF transceiver control and high-bandwidth USB communication (USB 3.2 Gen 1x1, up to 5\,Gbps).

% ┏━━━━━━━━━━━━━━━━━━━━━━━━━━━━━━━━━━━━━━━━━━━━━━━━━━━━━━━━━━━━━━━━━━━━━━━━━━┓
% ┃ Experimental Setup
% ┗━━━━━━━━━━━━━━━━━━━━━━━━━━━━━━━━━━━━━━━━━━━━━━━━━━━━━━━━━━━━━━━━━━━━━━━━━━┛
\subsection{Experimental Setup}
\label{sec:eval:setup}

We use a Raspberry Pi 4B~\cite{rpi4} equipped with a Broadcom BCM2711 SoC integrating four ARM Cortex-A72 cores as our target platform.
It runs on Raspberry Pi OS Lite (Linux kernel 6.12.25) with a command-line interface. 
The device operates at a fixed clock frequency of 600\,MHz without DVFS.
In all experiments, we use the prototype device to acquire per-core leakage signals with the center frequency set equal to the clock frequency and a sampling rate of 10\,MS/s. 
Each core is probed using high-resolution near-field probes---Langer MFA-R 0.2–6 and MFA-R 0.2–75 ($\Delta{=}300\,\um$)---based on the \CSMshort derived in~\secref{sec:eval:per-core}.
We also use the Riscure probe station to obtain the \CSMshort. 
The experimental setup is shown in~\figref{fig:setup}.

%% ┏━━━━━━━━━━━━━━━━━━━━━━━━━━━━━━━━━━━━━━━━━━━━━━━━━━━━━━━━━━━━━━━━━━━━━━━━━━┓
%% ┃ Per-Core Distinguishability
%% ┗━━━━━━━━━━━━━━━━━━━━━━━━━━━━━━━━━━━━━━━━━━━━━━━━━━━━━━━━━━━━━━━━━━━━━━━━━━┛
\subsection{Per-Core Distinguishability}
\label{sec:eval:per-core}

\begin{figure}[t]
    \centering%
    \includegraphics[width=0.65\linewidth]{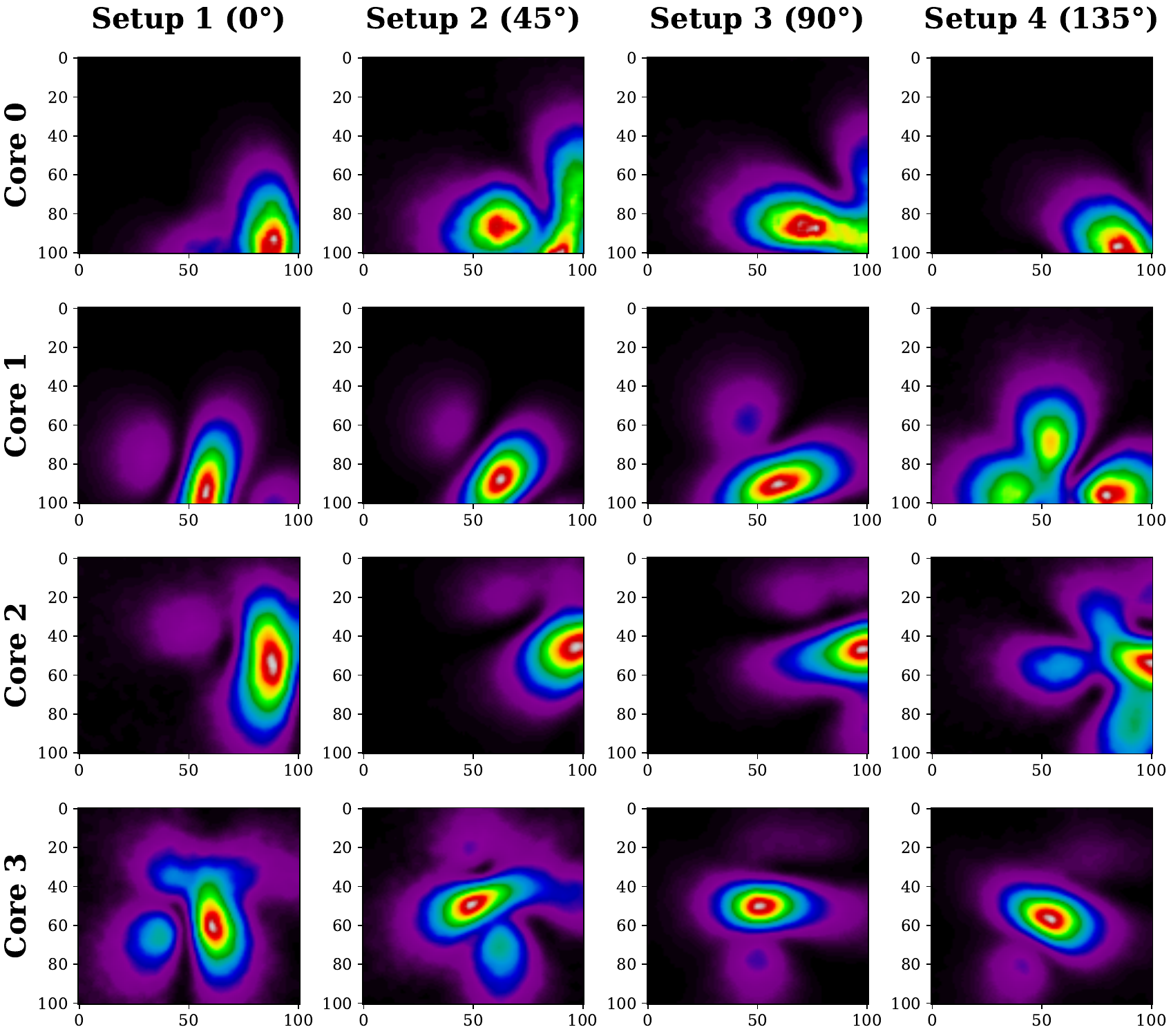}%
    \captionsetup{font=small}%
    \vspace{-0.15cm}%
    \caption{Leakage maps acquired from the four cores of the BCM2711 under four distinct probing setups (angles). Measurements were performed by scanning the surface with a 300\um-resolution coil positioned vertically to the die. }
    \vspace{-7pt}
    \label{fig:leakage-maps}%
    \vspace{-1.5ex}
\end{figure}

\begin{figure}[t]%
    \centering%
    \subfloat[\centering \CSM~(\CSMshort)\label{fig:csm}]{%
        \includegraphics[height=3cm]{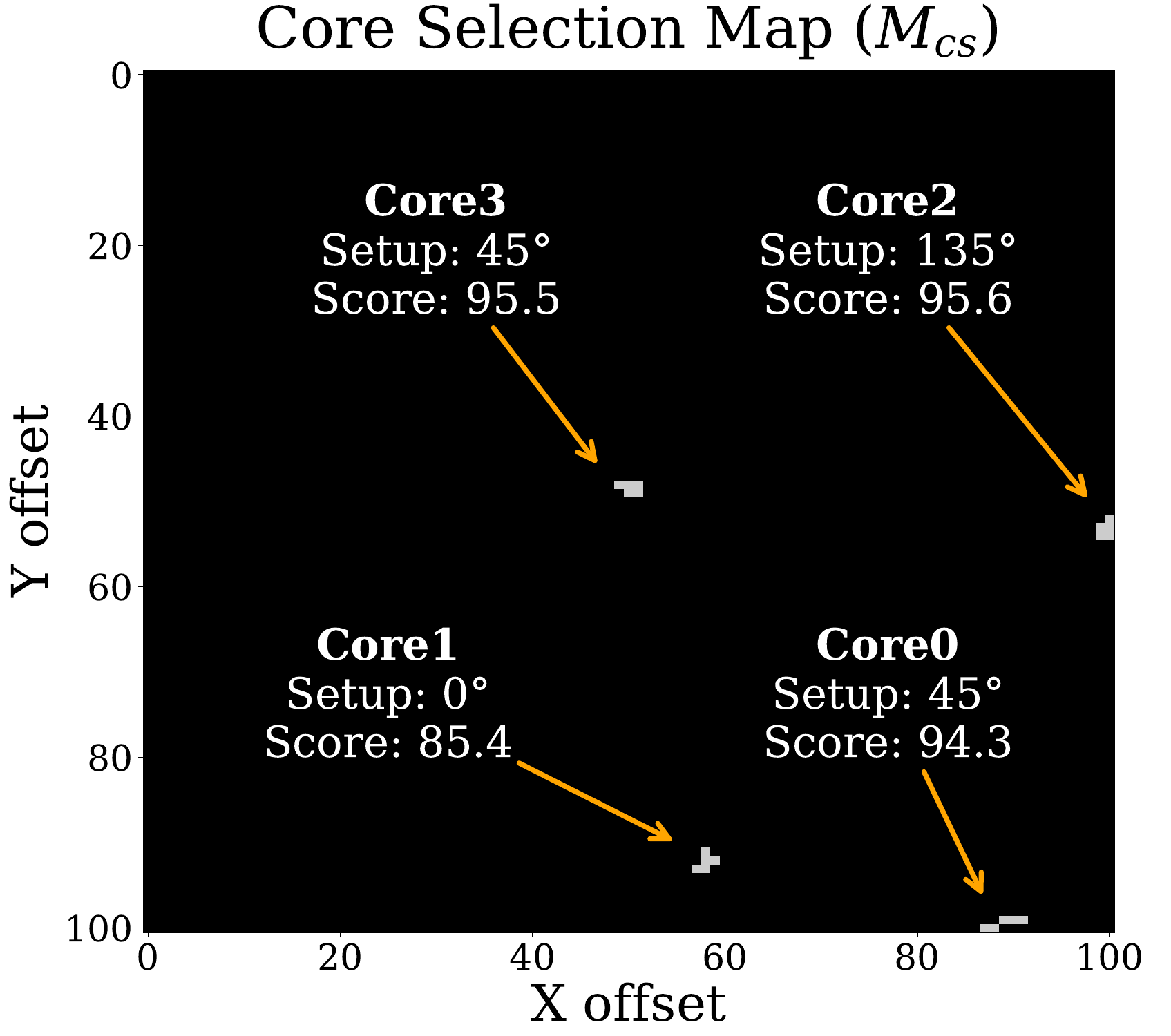}%
    }%
    \hspace{10pt}%
    \subfloat[\centering \IMX\label{fig:intf_matrix}]{%
        \includegraphics[height=3cm]{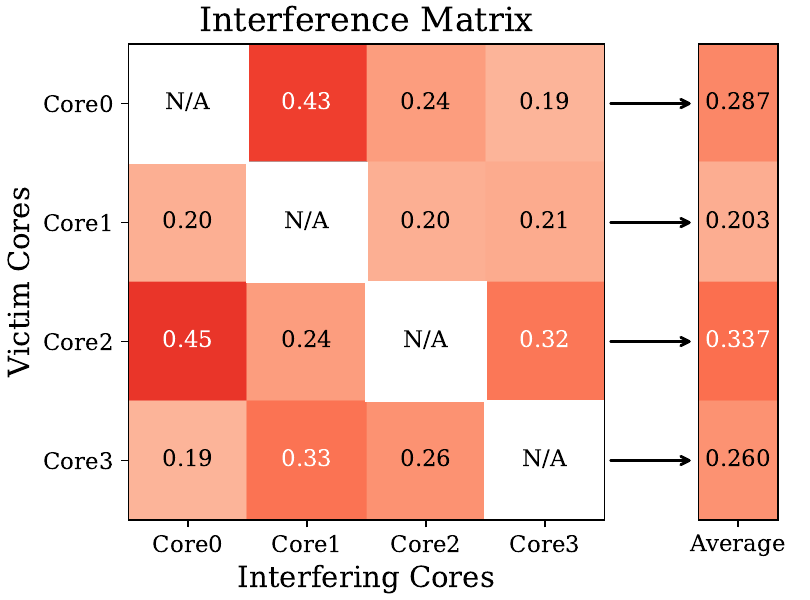}%
    }%
    \captionsetup{font=small}%
    \vspace{-0.25cm}
    \caption{\CSM and \IMX derived from \algref{alg:dpp} and \algref{alg:ais} using parameters $\alpha_1{=}15\%$, $\alpha_2{=}80\%$, $\beta{=}5$, $w{=}3$, and $N_{spectra}{=}1000$.}%
    \label{fig:csm_and_mat}%
    \vspace{-1.2ex}
\end{figure}

Using four different setups, we measure all four cores of the BCM2711 to obtain 16 per-core maps; 
these maps are shown in \figref{fig:leakage-maps} and constitute the \map{L}, from which the \CSMshort is derived using \algref{alg:dpp} (\figref{fig:csm}).
We also present the \IMX in~\figref{fig:intf_matrix}.
As described in~\secref{sec:method:AIS}, interference suppression proceeds in the order of cores least affected by neighboring activity---\texttt{Core}\,\texttt{1}$\rightarrow$\texttt{3}$\rightarrow$\texttt{0}$\rightarrow$\texttt{2}.
Acquiring the 16 leakage maps takes approximately 10 hours, while \algref{alg:dpp} completes in about 15 minutes (spatial granularity=$101{\by}101$).
As noted in~\secref{sec:method:DPP}, the \CSM is derived only once per device (or chip) type; thus, its duration does not hinder the feasibility.
\figref{fig:setup:probing} shows the high-precision multi-probing setup based on the \CSMshort.


%% ┏━━━━━━━━━━━━━━━━━━━━━━━━━━━━━━━━━━━━━━━━━━━━━━━━━━━━━━━━━━━━━━━━━━━━━━━━━━┓
%% ┃ Evaluation — Monitoring Capability
%% ┗━━━━━━━━━━━━━━━━━━━━━━━━━━━━━━━━━━━━━━━━━━━━━━━━━━━━━━━━━━━━━━━━━━━━━━━━━━┛
\subsection{Per-Core Monitoring: Leakage Exploitation}
\label{sec:eval:detection}

\begin{figure}[t]
    \centering%
    \includegraphics[width=0.9\linewidth]{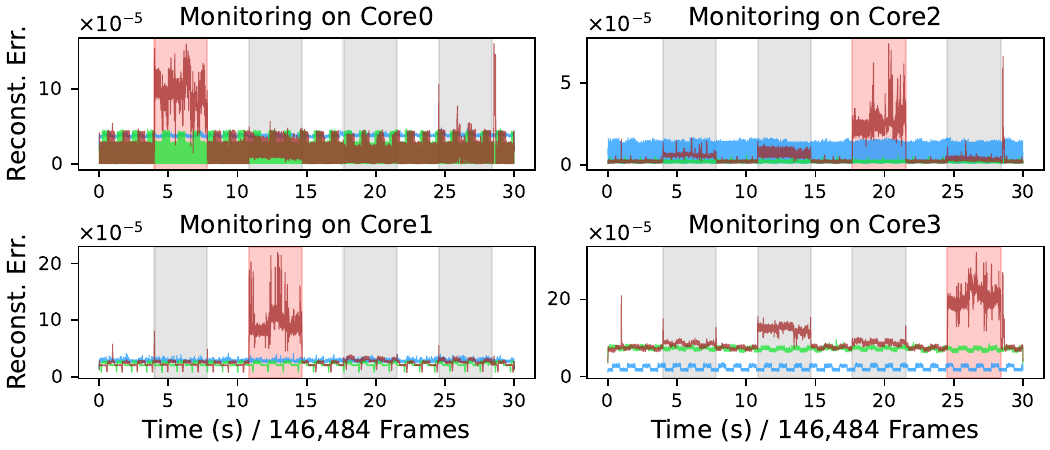}%
    \captionsetup{font=small}%
    \vspace{-0.15cm}%
    \caption{Per-core monitoring results on BCM2711 (SMA=15\,ms). \textcolor{green}{Green}:~benign behavior observed on the same core used during modeling (\textbf{\texttt{C0}:} \texttt{Susan}, \textbf{\texttt{C1}:} \texttt{Basicmath}, \textbf{\texttt{C2}:} MLP, \textbf{\texttt{C3}:} AES); \textcolor{blue}{Blue}:~benign behavior under a remapped core assignment (\textbf{\texttt{C0}:} AES, \textbf{\texttt{C1}:} MLP, \textbf{\texttt{C2}:} \texttt{Susan}, \textbf{\texttt{C3}:} \texttt{Basicmath}); \textcolor{red}{Red}:~anomalous behavior sequentially injected into \texttt{Core0} through \texttt{Core3} (PoC anomalous behavior: \texttt{find}).}%
    \label{fig:multi-core-result}%
    \vspace{-3ex}
\end{figure}

\begin{figure}[t]%
    \centering%
    \subfloat{%
        \includegraphics[height=3cm]{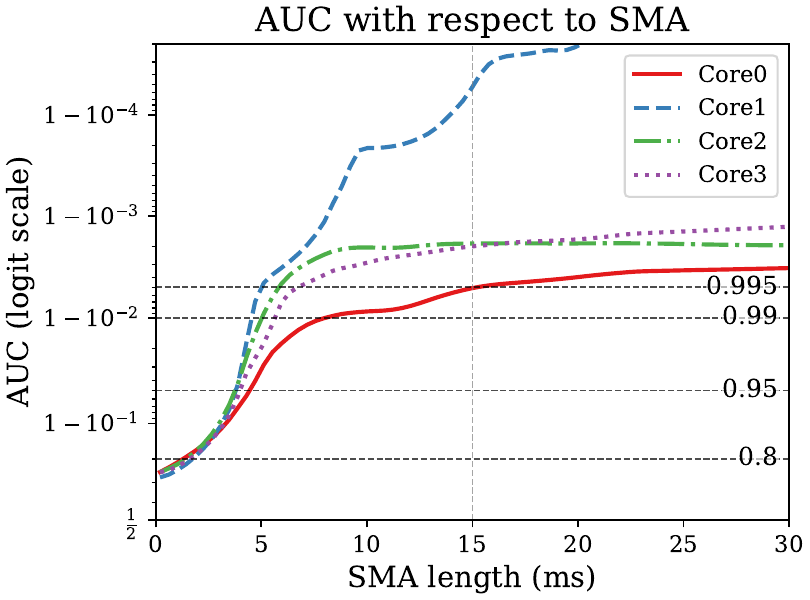}%
    }%
    \hspace{17pt}%
    \subfloat{%
        \includegraphics[height=3cm]{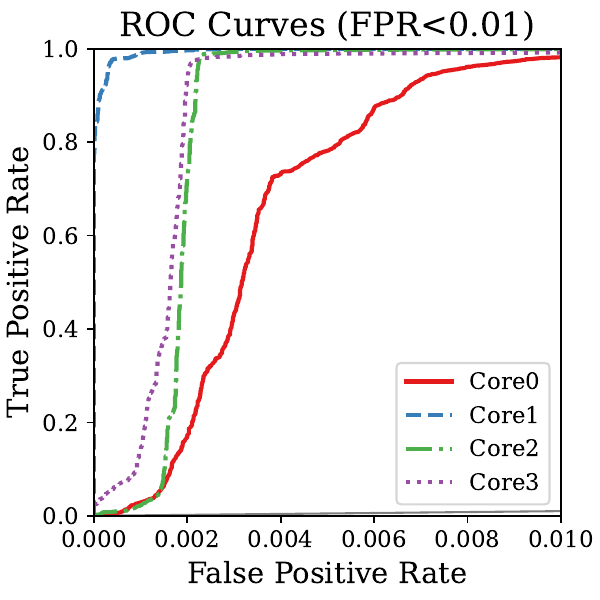}%
    }%
    \captionsetup{font=small}%
    \vspace{-0.15cm}
    \caption{Monitoring performance (AUC) with respect to SMA length and the corresponding ROC curves at SMA=15\,ms, where all cores achieve an AUC above 0.995.}%
    \label{fig:gran}%
    \vspace{-2ex}
\end{figure}

We define four software tasks as benign, taking into account the characteristics of embedded systems, and execute them in parallel on each core.
These include an MLP inference engine and an AES encryption module, reflecting trends in edge computing, as well as \texttt{basicmath} and \texttt{susan} from MiBench~\cite{wwc01:mibench}, which are widely adopted embedded benchmarks in side-channel monitoring literature.
We also model FFT frames of approximately 0.2\,ms in length using an autoencoder with a 2048–700–300–700–2048 architecture.
\figref{fig:multi-core-result} presents the results of sequentially triggering an undefined anomalous behavior on \texttt{Core0} through \texttt{Core3} while benign tasks are running concurrently on all cores.
To demonstrate core-agnostic monitoring, we also evaluate cases where the benign software is executed on cores different from those used during modeling (\textcolor{blue}{blue}).
\figref{fig:gran} illustrates the monitoring performance in terms of area under the curve (AUC) as a function of simple moving average (SMA) length and the receiver operating characteristic (ROC) curves at SMA=15\,ms, where all cores achieve an AUC above 0.995.

\section{Discussion}
\label{sec:discus}

Since \DPPshort requires high-precision physical positioning, the setup process tends to be both costly and cumbersome, and prone to spatial deviations.
This issue can be mitigated by employing custom-built high-precision probes with handcrafted coils~\cite{integration07:probe5, ICISSP24:probe1, ches14:probe2, poland13:probe3, wifs20:probe4}, as well as a harness designed based on \CSMshort.
In addition, repeatedly opening the device enclosure for each monitoring cycle is inefficient and may raise compliance issues---particularly in systems subject to electromagnetic compatibility (EMC)~\cite{iec61000} or military standards such as MIL-STD-461~\cite{milstd461g}. 
This issue can be mitigated by designing the target system with signal measurement ports at the enclosure level, considering EMI shielding when required;
the measurement setup can then be completed simply by connecting the instrument to external ports of the enclosure, assuming that internal probing is performed via a harness.

\section{Conclusion and Future Work}
\label{sec:conc}

This work, for the first time, unveils the EM leakage mechanisms of multicore architectures and establishes the feasibility of per-core leakage exploitation. 
Building on these findings, we develop a prototype that implements a side-channel monitoring method for multicore systems and validate its effectiveness on an off-the-shelf quad-core embedded platform.
By enabling per-core leakage exploitation---previously considered infeasible---these results are expected to facilitate broader EM-based side-channel analysis on modern multicore platforms.
Future work includes enhancing the proposed side-channel monitoring method, extending support to DVFS-capable systems, and integrating inference models into the FPGA fabric to achieve full end-to-end execution on the prototype.

\newpage
\bibliographystyle{ACM-Reference-Format}
\bibliography{
    bib/_IEEEabrv,
    bib/background,
    bib/device-ref,
    bib/prev_work_mcp,
    bib/prev_work_method,
    bib/prev_work_others,
    bib/probing-related,
    bib/standards,
    bib/term,
    bib/UNCLASSIFIED
}
\end{document}